\documentclass[a4paper,11pt]{article}
\usepackage{pos}

\title{The dwarf nova EX Draconis: a short review}
\ShortTitle{EX\,Dra: a short review}

\author*[a]{Raymundo Baptista}
\author[a,b]{Wagner Schlindwein}

\affiliation[a]{Departamento de F\'{i}sica, 
Universidade Federal de Santa Catarina - UFSC,\\
Campus Trindade, Florian\'{o}polis, SC, Brazil}

\affiliation[b]{Instituto Nacional de Pesquisas Espaciais
- INPE,\\ Avenida dos Astronautas, 1758, S\~ao Jos\'e dos
Campos-SP, Brazil}

\emailAdd{raybap@gmail.com}
\emailAdd{wagner.schlindwein@astro.ufsc.br}

\abstract{EX\,Draconis (EX\,Dra) is a long period dwarf nova showing
$\simeq 2$\,mag outburst which lasts for $\simeq 7$\,d and recur on a
timescale of $(20-30)$\,d. Its deep eclipses allows one to trace the
changes in surface brightness and radius of its accretion disk along
the outburst cycle and to perform critical tests of the predictions
of the thermal-viscous disk instability (DI) and the mass transfer
outburst (MTO) models proposed to explain dwarf nova outbursts.
The results of four critical tests are in clear contradiction with DI
while in good agreement with MTO expectations. Furthermore, the
observed variations in brightness and outer disk radius throughout
EX\,Dra outbursts are well described by the response of a high-viscosity
($\alpha= 3-4$) accretion disk to events in which the mass transfer rate
increases by factors of $\simeq 30$ for $\simeq 7$\,d, in line with MTO
expectations. We further argue that the old expectation of accretion disk
theory, $\alpha \lesssim 1$, seems unjustified and contradicts the
values derived from dwarf nova outburst decline timescales if they are
driven by MTO.}

\FullConference{The Golden Age of Cataclysmic Variables and Related Objects - VII (GOLDEN2025)\\
1-6 September, 2025\\
Mondello, Palermo, Italy\\}

\begin{document}
\maketitle

\section{Context}

Dwarf Novae (DN) are a sub-group of Cataclysmic Variables where a
non-strongly magnetic ($B\leq 10^5\,G$) white dwarf (WD) accretes matter
from a lower-mass, late-type donor companion (the secondary star) via an
accretion disk. DN show recurrent outbursts at days-months timescales,
in which the accretion disk increases in brightness by factors 20-100
during a few to several days \citep[e.g.,][]{warner2003}. Two models
were proposed in the 1970's to explain DN outbursts. In the thermal-viscous
Disk Instability (DI) model \citep{Osaki74,Cannizzo93,Lasota01,Hameury2020},
matter accumulates in a cool, low-viscosity
\footnote{here we adopt the prescription of \citet{ss} for
  the accretion disk viscosity, $\nu = \alpha c_s H$, where
  $\alpha$ is the non-dimensional viscosity parameter, $c_s$ is
  the local sound speed and $H$ is the disk vertical scaleheight.}
and unsteady disk during quiescence ($\alpha_c \sim 10^{-2}$, viscous
timescale longer than the outburst recurrence interval) and switches to
a hot, higher viscosity regime ($\alpha_h \sim 0.1-0.3$) during outbursts.
In the Mass Transfer Outburst (MTO) model \citep{bath,BathPringle81,SB2024},
the outburst is the viscous response of a disk with constant, high-viscosity
($\alpha\sim 1-3$, from the decline timescale of outbursting DN
\citep[e.g.,][]{mb83,warner2003}) to a burst of enhanced mass transfer
rate from the secondary star. Because of the permanent high-viscosity,
the disk is expected to be in a steady-state both in outburst and in
quiescence.

The widespread acceptance of DI as the correct explanation from the 1990's
onwards was largely affected by a crucial misconcept about the MTO, namely,
that an enhanced mass transfer stream would necessarily stop at disk rim,
leading to a significant increase in anisotropic emission from the bright
spot (BS) at outburst onset as well as preventing MTO to trigger inside-out
outbursts. The existence of inside-out outbursts and the lack of compelling
evidence for the increase in BS luminosity at outburst onset were taken as
strong evidence against MTO \citep[and references therein]{warner2003}.
However, numerical simulations of accretion disks show that when the gas
stream is denser than the outer disk material (the likelihood of
which increases with mass transfer rate \citep{bs22}), the gas stream
penetrates the disk and allows matter to be deposited in its inner regions
\citep{bisikalo98,makita,bisikalo05}, enabling inside-out outbursts while
leaving no enhanced BS emission footprint at outburst onset. Having this
in mind, the outcome of a MTO depends on the amplitude of the mass transfer
enhancement, $\Delta\dot{M}$ \citep{bem83}. In low $\Delta\dot{M}$ events,
there is no room for significant gas stream penetration and the burst of
matter is mostly deposited at disk rim, leading to outside-in outbursts
with both an increase in BS luminosity and a transient shrinking of the
disk at outburst onset (as a consequence of the large amount of gas of
low angular momentum added to the outer disk regions \citep[e.g.,][]
{LivioVerbunt1988}) before the subsequent disk expansion. In high
$\Delta\dot{M}$ events, the gas stream penetrates the disk and may bring
matter down to the circularization radius, leading to inside-out outbursts
with no BS luminosity increase and no significant disk shrinking at
outburst onset \citep{SB2024}.
There is no room for gas stream penetration in the extremely low-viscosity
($\alpha_c\simeq 10^{-2}$) and high density quiescent disks expected in
the DI framework \citep[e.g.,][]{bs22}.

EX~Dra is a relatively bright ($V\simeq 13.5$\,mag), long-period
($P_\mathrm{orb}= 5.04$\,h) and deeply eclipsing DN that shows
$\simeq 2$\,mag outbursts with a recurrence timescale of $\simeq (20-30)$\,d
\citep{Baptistaet2000,Courtet2020}. The trigonometric parallax distance
estimate ($241.2 \pm 1.3$\,pc \citep{Gaia2016,GaiaDR3}) is consistent
with that derived through photometric parallax ($290 \pm 80$\,pc
\citep{Baptistaet2000}).
Measurements of the radial velocity of the secondary star, the emission
lines (associated with the WD orbital movement), and the rotational
broadening of the secondary star lead to a purely spectroscopic model for
the binary \citep{Billingtonet1996,Fiedleret1997,SmithDhillon1998}, while
the ingress/egress phases of the WD and of the BS lead to a purely
photometric model of the binary \citep{Baptistaet2000,ShafterHolland2003}.
The photometric and spectroscopic models of the binary are consistent with
each other within the uncertainties, indicating that the masses and radii
of both stars, the orbital separation and orbital inclination are well
constrained. \citet{Harrisonet2004} derived a spectral type of K7 for
the secondary star, while \citet{Harrison2016} fitted its near-infrared
spectrum to obtain $T_2 = 4000$\,K, in agreement with the results of
\citet{ShafterHolland2003} and the CV evolutionary track of
\citet{knigge11}, which predicts $T_2= 3800\pm 200$\,K.

\begin{figure}
  \center
\includegraphics[scale=.6]{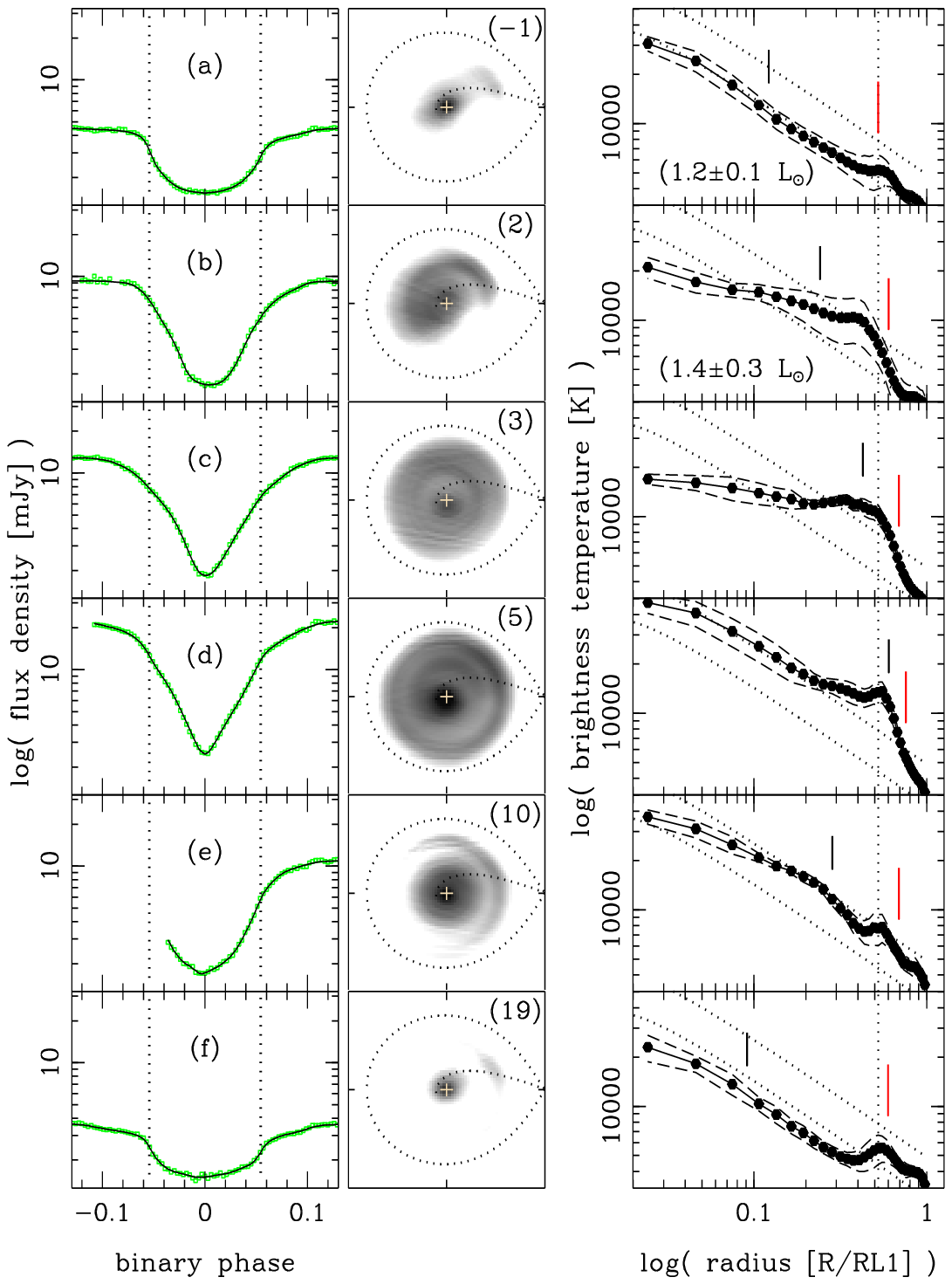}
\caption{Time-lapse eclipse mapping of the dwarf nova EX~Dra along its
  outburst cycle. Left-hand panels: Data (green dots) and model (solid
  line) light curves in (a) quiescent, (b) early rise, (c) late rise,
  (d) outburst maximum, (e) early decline, and (f) late decline stages.
  Vertical dotted lines mark ingress/egress phases of disk center.
  Middle panels: eclipse maps in a logarithmic grayscale; dark regions
  are brighter. Dotted lines show the primary Roche lobe and the
  ballistic trajectory of the gas from the secondary star; crosses mark
  the center of the disk. The secondary star is to the right of each panel;
  the stars and the accretion disk gas rotate counter-clockwise. The
  numbers in parenthesis indicate the time (in days) elapsed since outburst
  onset. Right-hand panels: azimuthally-averaged radial brightness
  temperature distributions for the eclipse maps in the middle panels.
  Dashed lines show the 1-$\sigma$ limit on the average temperature for
  a given radius. A dotted vertical line depicts the radial position of
  the BS in quiescence; vertical ticks mark the position of the outer
  edge of the disk (in red) and the radial position at which the disk
  temperature falls below 11000\,K (blue). Steady-state disc models
  for mass accretion rates of $\log$ \.{M}$= -8.0$ and $-9.0 \;M_\odot\;
  $yr$^{-1}$ are plotted as dotted lines for comparison. Numbers in
  parenthesis list the integrated disk luminosity. From \cite{bc01}.}
\label{time-lapse}
\end{figure}

The long orbital period (allowing eclipses to be sampled at a high phase
resolution), relatively high brightness (ensuring high signal-to-noise
light curves to be obtained), short outburst recurrence interval and deep
eclipses (allowing the use of eclipse mapping techniques to trace the
evolution of the disk surface brightness distribution along the
outburst cycle \citep{Horne1985,Baptista2016}) make EX\,Dra an ideal
environment to test the predictions of the DN outburst models. The
time-lapse eclipse mapping of EX\,Dra light curves throughout its outburst
cycle \citep[][see Fig.\,\ref{time-lapse}]{bc01} reveals the formation
of a one-armed spiral structure in the disk at the early stages of the
outburst (Fig.\,\ref{time-lapse}b) and that the disk expands during the
rise phase until it fills most of the primary Roche lobe at maximum light
(Fig.\,\ref{time-lapse}d). During the decline phase, the disk becomes
progressively fainter (Fig.\,\ref{time-lapse}e) until only a small bright
region around the WD is left at minimum light (Fig.\,\ref{time-lapse}f).
Analysis of the radial brightness temperature distributions indicates
that the disk appears to be in a steady-state during quiescence
and at outburst maximum, but not during the intermediate stages.
As a general trend, the mass accretion rate in the outer regions is
larger than that in the inner disk in the ascending branch, while the
opposite occurs during the descending branch. Fitting opaque steady disk
models to radial temperature distributions allows one to estimate accretion
rates of $\dot{M}= 10^{-7.7 \pm 0.3}$\,M$_\odot$\,yr$^{-1}$ ($1.6_{-0.8}^{+1.6}
\times 10^{18}$\,g\,s$^{-1}$) at outburst maximum and $\dot{M}=
10^{-9.1 \pm 0.3}$\,M$_\odot$\,yr$^{-1}$ ($5.0_{-2.5}^{+5.0} \times
10^{16}$\,g\,s$^{-1}$) in quiescence \cite{bc01}.
%

Here we report critical tests of the predictions of the DI and MTO
models against the EX\,Dra eclipse mapping results 
(Sect.\,\ref{tests}), we show that the variations in brightness and
disk radius throughout the EX\,Dra outburst can be well explained in the
framework of the MTO model (Sect.\,\ref{mto-model}), and we discuss the
implications of our results in the context of DN outbursts and accretion
disk theory (Sect.\,\ref{discuss}). The results are summarized in
Sect.\,\ref{summary}.

\section{Testing outburst models with EX~Dra eclipse mapping results}
\label{tests}

Observations of eclipsing dwarf novae throughout their ourburst cycle
provide a key opportunity to test the predictions and to discriminate
between the DI and MTO explanations \citep{bap12}. In particular, the
critical tests are those the results of which are consistent with
predictions of one of the models, but inconsistent with those of the other.

\subsection{The quiescent disk}

DI predicts that quiescent DN harbor low-viscosity ($\alpha_c \leq
0.05$ \cite{hameury98}), unsteady accretion disks with a flat radial
temperature distribution and a slow viscous response to eventual
changes in mass transfer rate, while MTO predicts high-viscosity
steady-state quiescent disks, with a fast response to changes in mass
transfer rate. The differences in prediction are very significant.
For EX\,Dra, the estimated viscous timescale of a quiescent DI disk
is $t_\mathrm{visc}(\mathrm{DI})\geq 60$\,d, which is much longer than its
$(20-30)$\,d outburst recurrence time -- indicating that there is not
enough time to reach a state-state during quiescence \citep{SB2024}.
However, the eclipse mapping analysis of \citet{bc01} shows
that both the radial temperature distribution at minimum light
(Fig.\,\ref{time-lapse}f) and in quiescence (Fig.\,\ref{time-lapse}a)
closely follow the $T\propto R^{-3/4}$ law of opaque steady-state discs --
respectively with mass accretion rates of $\dot{M}(f)= (2.3\pm 0.3)\times
10^{16}$\,g\,s$^{-1}$ and $\dot{M}(a)= (5.3\pm 0.6)\times 10^{16}$\,g\,s$^{-1}$
-- and that the transition between these two steady-states occur in a viscous
timescale of only $t_\mathrm{visc}(\mathrm{quies})= 1.5$\,d, implying a
quiescent viscosity parameter of $\alpha \simeq 2$, in line with the high
viscosity range inferred in Sect.\,3. These results are in clear
contradiction with DI while in good agreement with MTO expectations.

\subsection{Accumulated mass versus accreted mass}

In the DI framework, the maximum mass accumulated in the disk during
quiescence, $\Delta M_q\mathrm{(max)} \\ = \dot{M}_q \Delta t_q$ (derived
under the assumption that all mass being transferred in quiescent is
stored in the disk), is accreted onto the WD during outburst, 
$\Delta M_o= \dot{M}_o \Delta t_o \leq \Delta M_q\mathrm{(max)}$,
at a predicted maximum outburst mass accretion rate of,
\begin{equation}
  \dot{M}_o\mathrm{(max)} = \frac{\Delta t_q}{\Delta t_o} \dot{M}_q \,\, ,
  \label{eq:mdot-out}
\end{equation}
where $\Delta t_q$, $\Delta t_o$ and $\dot{M}_q$, $\dot{M}_o$ are,
respectively, the durations of the quiescence and of the outburst phases
and the mass accretion rates in quiescence and in outburst.
For EX\,Dra, $\Delta t_q\simeq (13-23)$\,d, $\Delta t_o\simeq 7$\,d
\citep{Baptistaet2000}, and one expects $\dot{M}_o\mathrm{(max)} \simeq
(2-3)\,\dot{M}_q \simeq (1.0-1.5)\times 10^{17}$\,g\,s$^{-1}$.
However, the inferred mass accretion rate at outburst maximum
($1.6\times 10^{18}$\,g\,s$^{-1}$ \citep{bc01}) is one order of magnitude
larger than the DI prediction, implying that the mass accreted during
outburst is also one order of magnitude larger than the maximum mass that
could have accumulated in the disk during the previous quiescent phase.
It is not possible to reconcile DI and the observations with such
significant discrepancy, and one is lead to the conclusion that the
outbursts of EX\,Dra must involve a significant increase in mass transfer
rate during the outburst itself -- in accordance with MTO expectations.

\subsection{Enhanced stream emission at early rise}
\label{rise}

\begin{figure}
  \center
  \includegraphics[width=0.55\textwidth]{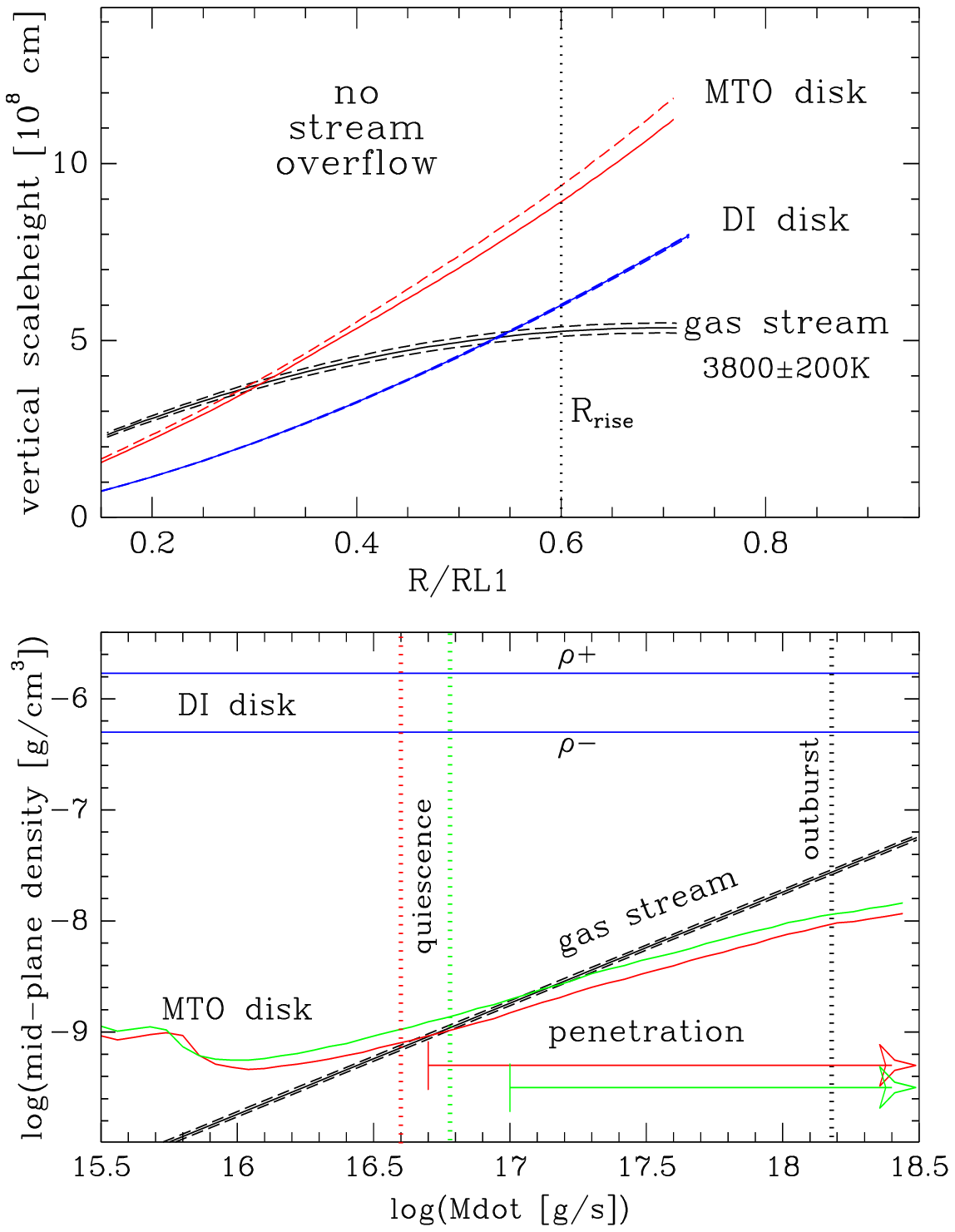}
  \caption{Top: Comparison of the vertical scaleheights of the disk and gas
  stream for EX\,Dra. Radial runs of the disk scaleheight are shown for
  DI (blue) and MTO (red, solid line for $\alpha=4$ and dashed line for
  $\alpha=3$) disk models. The vertical scaleheight of the gas stream is
  shown for $T_2= 3800\pm 200$\,K; dashed lines depict the corresponding
  1-$\sigma$ limits. A vertical dotted line marks the outer disk radius
  at early rise, $R_d=0.6\,R_\mathrm{L1}$ \citep{bc01}.
  Bottom: Disk and gas stream midplane densities as a function of mass
  transfer rate for a disk radius of $R_\mathrm{d}= 0.6\,R_{L1}$. The red
  (green) lines show the MTO disk midplane densities for $\alpha=4$
  ($\alpha=3$), while blue lines show the range of possible DI disk
  midplane densities. Black lines show the gas stream midplane densities
  for $T_2= 3800\pm 200$\,K. Vertical dotted lines mark the inferred mass
  transfer rates in quiescence (in red for $\alpha=4$ and in green for
  $\alpha=3$) and in outburst.
  \label{stream}}
\end{figure}

The early rise eclipse map (Fig.\,\ref{time-lapse}b) shows evidence of
enhanced gas stream emission beyond impact at disk rim, suggesting the
occurrence of gas stream overflow or penetration on that occasion
(the reader is referred to \citet{bs22} for a comprehensive discussion
on the possibilities of gas stream overflow and/or penetration in DN
accretion disks). The upper panel of Fig.\,\ref{stream} compares the
expected vertical scaleheight of EX\,Dra early rise DI and MTO
($\alpha=3$ and 4) disk models with the predicted vertical scaleheight of
the gas stream for a secondary star of temperature T$_2= 3800\pm 200$\,K
as a function of radius in units of the distance from WD to the inner
lagrangian point L1. No stream overflow is possible at this outburst stage,
as the disk scaleheights at disk rim ($R_d=0.6\,R_\mathrm{L1}$) are larger
than the gas stream scaleheight at least at the 3-$\sigma$ confidence level.
The lower panel of Fig.\,\ref{stream} compares the midplane densities of
the EX\,Dra early rise DI and MTO disks models ($\rho_d$) with the midplane
density of the gas stream ($\rho_s$), at disk rim, as a function of
mass transfer rate. Stream penetration ($\rho_s>\rho_d$) onto a DI disk
might only occur at extremely large mass transfer rates, $\dot{M}_2 >
10^{19}$\,g\,s$^{-1}$, far into the hot branch where the EX\,Dra accretion
disk becomes stable against DI-driven outbursts -- in marked contrast
with its outbursting nature.
On the other hand, stream penetration onto a MTO disk is much more
plausible; it starts to occur at mass transfer rates slightly above the
inferred quiescent mass transfer rate.
Whereas the enhanced gas stream emission at early rise cannot be
explained in the DI framework, it is the natural outcome of an increase
in mass transfer rate onto a high-viscosity disk -- as predicted
for the EX\,Dra early outburst stage in the MTO framework.
%

\subsection{Enhanced mass-transfer at early rise: cause or consequence?}

The results of Sect.\,\ref{rise} are indicative of an enhanced mass transfer
rate at this early outburst stage. The integrated disc luminosity at early
rise ($1.4 \pm 0.3\,L_\odot$, Fig.\,\ref{time-lapse}b) is comparable to that
in quiescence ($1.2 \pm 0.1\,L_\odot$, Fig.\,\ref{time-lapse}a) and is not
enough to support the idea that the enhanced mass transfer rate could be
triggered by an increased irradiation of the secondary star by the
accretion disk. This led \citet{bap12} to the conclusion that the observed
enhanced mass transfer rate at early rise is not a consequence of the
ongoing outburst, but its cause -- suggesting that the outbursts of EX\,Dra
are powered by events of enhanced mass transfer, as expected by MTO.

\section{Mass-transfer outburst modelling}
\label{mto-model}

None of the critical tests performed in Sect.\,\ref{tests} is consistent
with DI; all are in good agreement with MTO predictions. Given that the
observational evidence largely favors MTO, we may ask the question:
can MTO provide a satisfactory explanation of the disk radius and
brightness changes throughout the EX\,Dra outburst?

\citet{SB2024} developed an MTO code to simulate the response of an
accretion disk to events of enhanced mass transfer, which includes a
mass transfer event of smooth shape, mass deposition taking into
account gas stream penetration
\footnote{The gas stream penetration radius ($R_p$) is taken as the
  radius where the midplane gas stream density equals the midplane disk
  density. There is gas stream penetration when $R_p<R_d$ \citep{bs22}.}
, and disc emission with either blackbody or gray atmosphere approximation.
The smooth shape of the mass transfer event results in continuous changes
in disk radius along the outburst and leads to a delay between the
increase in $\dot{M}$ in the outer regions and the accretion onto the WD,
akin to the well known UV-delay effect seen in the initial outburst stages
of several DN \citep[e.g.,][]{shl2005}. Gas stream penetration allows for
inside-out outbursts with minimal disk shrinkage at outburst onset, and
gray atmospheres allow for optically thin outer disk regions.

\begin{figure}
  \center
\includegraphics[width=0.5\columnwidth]{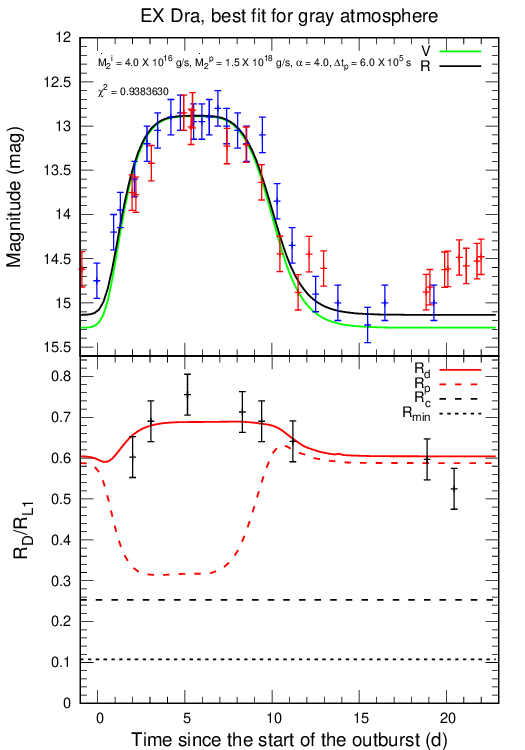}
\caption{Top: Average $V$- (blue points) and $R$-band (red points) EX\,Dra
  outburst light curves with respect to the time since outburst onset,
  together with corresponding best-fit MTO model light curves for gray
  atmosphere local emission. The input parameters for this model are listed
  together with the $\chi^2$ value of the fit. Bottom: Changes in disk radius
  as a function of time from outburst onset. The disk radius measurements by
  \citep{bc01} are shown as points with error bars in comparison to the
  MTO model outer disk radius (solid red line) and the stream penetration
  radius (dashed red line). Black dashed and dotted lines depict the
  circularization radius and the radius of shortest stream distance from
  the WD, respectively.
 \label{brilho-raio}}
\end{figure}

This MTO simulation code was used to simultanously model the observed
variations in brightness at the $V$- and $R$-bands and in outer disk
radius throughout the outburt cycle of EX\,Dra \citep{SB2024}.
Observations made by amateur astronomers from AAVSO and VSNET were
used to obtain a representative $V$-band light curve of EX\,Dra by
combining measurements covering 14 outburst cycles, aligned according
to the start of the rise to maximum. Only outbursts with amplitude and
duration similar to those covered by the observations of \citet{bc01}
were included. The data set was sliced in groups of 10 points and a
median magnitude was computed for each group. The results are shown as
blue symbols with error bars in the upper panel of Fig.\,\ref{brilho-raio}.
Red symbols indicate $R$-band out-of-eclipse magnitudes measured by
\citet{bc01}. These are typical type B (inside-out) outbursts
\citep{Smak1984,warner2003}, with comparable rise and decline timescales.
The lower panel of Fig.~\ref{brilho-raio} shows the evolution of the
outer disk radius as measured by \citet{bc01}. There is no evidence of
a decrease in outer disk radius at outburst onset.

A large grid of MTO models was built, covering a range of values for the
4 input parameters: quiescent mass transfer rate, $\dot{M}_2^i$, mass transfer
rate at outburst maximum, $\dot{M}_2^p$, event duration, $\Delta t_p$, and
viscosity parameter, $\alpha$. The value of $\dot{M}_2^i$ is determined by
the magnitude in quiescence, and that of $\dot{M}_2^p$ is constrained by
the magnitude during the plateau. The value of $\Delta t_p$ is derived
from the full-width-at-half-maximum of the outburst and the value of
$\alpha$ is defined by the outburst decline timescale. The best fit
outburst model to the observations was found by calculating the $\chi^2$
of the fit for each grid model. Solid lines in Fig.\,\ref{brilho-raio}
show the best-fit model for the case of gray atmosphere local emission.
The MTO model predicts significant gas stream penetration during outburst
(red dashed line in the lower panel of Fig.\,\ref{brilho-raio}), which
eliminates the marked initial reduction in disk radius predicted by the
simulations of \citet{LivioVerbunt1988} and \citet{IchikawaOsaki1992}
and leads to an outer disk radius variation consistent with the EX\,Dra
observations.

The observed variations in brightness and outer disk radius throughout
the outbursts of EX~Dra are well described ($\chi^2\simeq 1$) by the
response of a high-viscosity accretion disk to an event of enhanced mass
transfer rate, in accordance with MTO predictions. The best-fit MTO models
indicate viscosity parameters in the range $\alpha= 3$ (blackbody local
emission) to $\alpha=4$ (gray atmosphere local emission) and a mass
transfer event of width $\Delta t_p= 6\times 10^5\,\mathrm{s} \simeq 7$\,d.
The values of the quiescence and outburst maximum mass transfer rates
inferred from the MTO model are consistent with those derived by
\citet{bc01} at the 1-$\sigma$ limit \citep{SB2024}.

\section{Is alpha larger than unity unphysical?}
\label{discuss}

In their effort to overcome the absense of a theory of viscosity in
differentially rotating fluid disks, \citet{ss,ss76} parametrized the
kinematic viscosity in terms of the known quantities $c_s$ and $H$,
and transferred the ignorance on the unknown viscosity mechanism to
the non-dymensional $\alpha$ parameter,
\begin{equation}
\alpha = \frac{v_t l_t}{c_s H} + \frac{B^2}{4\pi \rho c_s^2} \, ,
\label{eq-alpha}
\end{equation}
where
$v_t$ and $l_t$ are respectively the turbulent velocity and the turbulent
mixing length (in case of hydrodynamic turbulence), $B^2/8\pi$ is the
energy density of the chaotic magnetic field and $\rho c_s^2/2$ is the
thermal energy density of the matter in the disk (in case of
magneto-hydrodynamic turbulence \citep[e.g.,][]{bh91,hb91}).
The theoretical expectation $\alpha \lesssim 1$ is based on the
assumptions that (i) the turbulent mixing length can not exceed the
disk vertical scaleheight ($l_t \leq H$) and (ii) the turbulence must be
subsonic ($v_t \lesssim c_s$) otherwise the turbulent motions would
probably be termalized by shocks \citep{ss,acpower}.

\citet{acpower} noted that $v_t > c_s$ could occur in disk regions
where some physical input continually feeds a supersonic turbulence.
\citet{2019.NewAstr.70.7} remarked that both the disk mass inflow and the
rate of heating by viscous dissipation depend linearly on the value of
$\alpha$ and that, while a larger $\alpha$ leads to a larger amount
of energy dissipation (presumably through shocks), it also leads to a
larger accretion rate which can provide the necessary larger energy to be
dissipated. Hence, they concluded that the $v_t \lesssim c_s$ assumption
is not justified on energy arguments. In addition, \citet{SB2024} argue
that there is no reason to exclude the possibility that shocks in
trans-sonic or supersonic turbulence may just provide the dissipative
process required for viscous heating in accretion disks.

The $l_t \leq H$ assumption is based on the idea that turbulence is
isotropic. However, in an accretion disk angular momentum exchange
occur in the radial direction, strong stretching of turbulent eddies or
magnetic field lines occur in the azimuthal direction, and most of the
energy resulting from viscous dissipation is transported in the vertical
direction. It makes not much sense to believe that (hydrodynamical or
magneto-hydrodynamical) turbulence in such an anisotropic environment
should be isotropic \citep{SB2024}.
Indeed, numerical simulations of the magneto-rotational instability
show how an initially isotropic magnetic field rapidly becomes chaotic,
anisotropic and significantly stretched in the radial direction, with
the amount of field stretching being limited mostly by the radial width
of the shearing box \citep[][see their Fig.\,9]{hb91}. In addition,
the radial stretching of the magnetic field lines allows exchange of
angular momentum over radial distances which greatly exceed the linear
wavelength of the instability (which will be present as long as the
minimum unstable wavelength do not exceed the disc thickness), thus
indicating that $l_t > H$.

Given that both $l_t > H$ and $v_t > c_s$ are valid possibilities, the
theoretical expectation $\alpha \lesssim 1$ seems unjustified and there
is no {\em a priori} reason to consider that $\alpha > 1$ is unphysical
\citep{SB2024}.

Since the definition of $\alpha$ in Eq.~\ref{eq-alpha} is
independent of both the isotropic and the subsonic assumptions,
the parametrization of \citet{ss,ss76} is still applicable if
these two assumptions are dropped.
In the case of magneto-hydrodynamic turbulence, we may define
the components of the Alfv\'en velocity as $v_{Ai}^2= B_i^2/4\pi\rho$.
For an anisotropic magnetic field, it is possible (and likely
\citep{hb91}) to have $v_{A\phi}^2 , v_{AR}^2 >
c_s^2 > v_{Az}^2$ and, therefore, $\alpha = v_A^2/c_s^2 > 1$.
Similarly, in the case of anisotropic hydrodynamic turbulence,
the turbulent mixing length is allowed to be different for each
direction, $l_\phi \neq l_R \neq l_z$ (from the dynamics of an
accretion disk one might expect $l_\phi > l_R > l_z$). Therefore,
it is possible (and likely) to have $l_\phi,l_R > H > l_z$ and
$\alpha>1$ as well, even in the case of subsonic turbulence.
The accretion disk theoreticians are encouraged to explore
this still untouched domain.

\section{Summary}
\label{summary}

This short review of EX\,Dra can be summarized as follows:

\begin{itemize}

\item EX\,Dra is an eclipsing version of SS\,Cyg (the {\em bona fide}
  long-period DN), a gift provided by nature to allow us to test the
  current theories about DN outbursts.

\item Time-lapse eclipse mapping of EX\,Dra throughout its outburst allow
  us to perform four critical tests of the DN outburst theories. The results
  of all four tests are in clear contradiction with DI while in good
  agreement with MTO expectations.

\item The observed variations in brightness and outer disk radius throughout
  EX\,Dra outbursts are well described ($\chi^2\simeq 1$) by the response of
  a high-viscosity ($\alpha= 3-4$) accretion disk to events in which the
  mass transfer rate increases by factors of $\simeq 30$ for $\simeq 7$\,d,
  in line with MTO expectations.

\item EX\,Dra is no anomaly; it teams with V2051\,Oph, HT\,Cas, 
  V4140\,Sgr, EX\,Hya, YZ\,LMi, and possibly SS\,Cyg and OY\,Car in an
  increasing group of DN the outbursts of which seems to be powered
  by MTO instead of by DI.

\item There is no physical support for the assumptions that (i) the
  turbulent mixing length has to be lower than the disk vertical
  scaleheight and (ii) turbulence must be subsonic.
  The theoretical expectation $\alpha \lesssim 1$ is unjustified.
  
\item If outbursts in DN and soft x-ray transients are viscous events
  (instead of thermal-viscous instability events), than the viscosity
  parameters inferred from their outburst decline timescales
  are systematically larger than unity and the theoretical expectation
  $\alpha \lesssim 1$ is inconsistent with observations.

\end{itemize}

\acknowledgments R. Baptista acknowledges finantial support from CNPq
grant 421034/2023-8. This study was financed in part by the
Coordena\c{c}\~ao de Aperfei\c{c}oamento de Pessoal de N\'{i}vel
Superior - Brasil (CAPES) - Finance Code 001.

\bigskip
\bigskip
\noindent {\bf DISCUSSION}

\bigskip
\noindent {\bf ALLEN SHAFTER:} What instability in the secondary star
results in the modified mass transfer?

\bigskip
\noindent {\bf RAYMUNDO BAPTISTA:} The original idea of an instability
in the atmosphere of the secondary star \cite{bath,BathPringle81} was
abandoned a long time ago. Currently, the most promising explanation
for the sudden changes in mass transfer rate required by MTO involves
starspots moving in and out of the inner lagrangian point L1. 
\citet{lp94} and \citet{kc98} pointed out that the passage of starspots
in front of L1 can significantly reduce $\dot{M}_2$, leading 
\citet{hl14} to suggest that events of enhanced mass transfer may
occur when there are no starspots transiting L1.

\end{document}